\def\e{{\rm e}}
\def\del{\partial}
\def\half{{1\over2}}

\def\vev#1{\langle #1 \rangle}
\def\del{\partial}
\def\dslash{\del\kern-0.55em\raise 0.14ex\hbox{/}}

\def\rough#1{\raise.3ex\hbox{$#1$\kern-.75em\lower1ex\hbox{$\sim$}}}
\newcommand{\PRD}[3]{Phys. Rev. {\bf D{#1}}, {#2} (19{#3})} 
\newcommand{\PRDM}[3]{Phys. Rev. {\bf D{#1}}, {#2} (20{#3})} 
 
\newcommand{\NPB}[3]{Nucl. Phys. {\bf B{#1}}, {#2} (19{#3})} 
\newcommand{\NPBM}[3]{Nucl. Phys. {\bf B{#1}} (20{#2}) {#3}}
\newcommand{\PLB}[3]{Phys. Lett. {\bf B{#1}}, {#2} (19{#3})}

\newcommand{\ANN}[3]{Ann. Phys. (N.Y.) {\bf {#1}}, {#2} (19{#3})}

\newcommand{\hmu}{\hat\mu}

\documentclass[12pt]{article}
\begin{document}
%%%%%%
% Definition of title page:
\title{Effect of Bare Mass on the Hosotani Mechanism}
\author{Kazunori Takenaga \vspace{1cm}$^{}$\thanks 
{email address: takenaga@het.phys.sci.osaka-u.ac.jp}\\ 
%%%
%\it {School of Theoretical Physics,}\\ 
%%%
{\it Department of Physics, Osaka University,}\\
{\it  Toyonaka, Osaka 560-0043, Japan}}
\date{} %optional
\maketitle
%%%%%
\baselineskip=18pt
%%%%
%\baselineskip=0.5 truein plus 2pt minus 1pt
%%%%%
\vskip 2cm
\begin{abstract}
%\baselineskip=0.5 truein plus 2pt minus 1pt
%%%%%%%%ABSTRACT %%%%%%%
It is pointed out that the existence of bare mass terms for matter fields
changes gauge symmetry patterns through the Hosotani mechanism. As a 
demonstration, we study an $SU(2)$ gauge model with massive adjoint fermions
defined on $M^4\otimes S^1$. It turns out that the vacuum 
structure changes 
at certain critical values of $mL$, where $m~(L)$ stands for the bare 
mass (the circumference of $S^1$). The gauge symmetry breaking 
patterns are different from models with  
massless adjoint fermions. We also consider a 
supersymmmetric $SU(2)$ gauge model with adjoint hypermultiplets, in which 
the supersymmetry is broken by bare mass terms for the gaugino and 
squark fields instead of the Scherk-Schwarz mechanism.
%%%%%%%%%%%
%We compare the gauge symmetry breaking patterns of the model with those of 
%the one with the Scherk-Schwarz mechanism of supersymmetry breaking.   
%%%%%%%%%%%
\end{abstract}
%%%%%preprint number %%%%%%%
\vskip 2cm
\begin{flushleft}
%%
%DIAS-STP-03-05\\
%%%
OU-HET-447/2003\\
%%%
May 2003
%{\tt hep-th xxxxxxx}\\
\end{flushleft}
%%%%%%% PACS number(s) %%
\begin{center}
%To appear in Phys. Rev. {\bf D}
%PACS number(s): 11.15.Ex, 11.25.Mj, 12.60.Jv
\end{center}
%%%%%%%%%%
\addtolength{\parindent}{2pt}
\newpage
%%%%%%%%%%%%%%%%%%%%%%%%%%
%%%%%% BODY %%%%%%
%%%%%% introduction %%%%%%
\section{Introduction}
The Hosotani mechanism is one of the most important dynamical phenomena 
when one considers physics with compactified extra 
dimensions~\cite{hosotani}\cite{hosotanib}. The component 
gauge field for the compactified direction becomes 
dynamical degrees of freedom and can develop vacuum expectation values, which
are related to phases of the Wilson line integrals along the
compactified direction. And the Wilson line is an order parameter
for the gauge symmetry breaking. Quantum corrections in
the extra dimension are crucial for the mechanism and gauge symmetry can be 
broken dynamically, reflecting the topology of the extra dimension. 
\par
%%%%%%%%%
It is important to note that the mechanism is essentially governed 
by infrared physics.  In order to study vacuum structures of the 
theory, one usually studies the effective potential for the Wilson line 
phases. The Wilson line is a global quantity, so that the effective potential is 
free from ultraviolet effects because they are local.
This suggests that massive particles, at first glance, do not 
contribute to the effective potential for the phases. Thus, the gauge 
symmetry breaking patterns seem not to be affected by the massive particle.
\par
%%%%%%%%%%%%%%%%
In this paper we demonstrate that massive
particles, on the contrary, can affect the gauge symmetry breaking 
patterns through the Hosotani mechanism. In fact we will show that in gauge
models with massive fermions, the gauge symmetry breaking patterns change
at certain critical values of the bare mass for the fermion.
This is never observed in the case of massless fermions.
\par
%%%%%%%%%%%%%
We also study supersymmetric gauge models, in which the supersymmetry 
is broken explicitly by bare mass terms, instead of imposing the twisted 
boundary condition for the $S^1$ direction by 
Scherk and Schwarz~\cite{ss}.
We introduce a bare mass term for the gaugino(squark) field in 
a vector(hyper)multiplet. We will show that, depending on the relative
magnitude between the two mass parameters, the gauge symmetry 
breaking patterns become different from those by the model with 
Scherk-Schwarz mechanism of supersymmetry breaking.
\par
%%%%%%%%%%%%%
Since the full analysis is beyond the scope of this paper, we restrict our 
consideration to a simple $SU(2)$ (supersymmetric) gauge model 
with $N_f$ massive 
adjoint fermions (hypermultiplets) defined on $M^4\otimes S^1$, where 
$M^4(S^1)$ stands for four-dimensional Minkowski space-time(a circle with the 
circumference being $L$). And we are interested in the effects of 
bare mass terms for matter fields on the gauge symmetry breaking patterns.
%%%%%%%%%%%%%%%
\section{Gauge model with massive adjoint matter}
Let us start with the nonsupersymmetric $SU(2)$ gauge model
with $N_f$ massive adjoint fermions defined on $M^4\otimes S^1$. 
Following the standard procedure, the effective potential for the 
constant background gauge field 
$gL\vev{A_y}=\mbox{diag}(\theta_1, \theta_2)=\mbox{diag}(\theta, -\theta)$ 
is given by~\cite{pomarol}\cite{hosotanib} 
%%%%%%%%%%%%%%
\begin{eqnarray} 
V_{eff}&=&-(5-2){{\Gamma({5\over 2})}\over {\pi^{{5\over 2}}}}
{1\over L^5}
\sum_{i, j=1}^2\sum_{n=1}^{\infty}{1\over n^5}\cos
\left(n(\theta_i -\theta_j)\right)
\nonumber\\
&+&
\left(N_f2^{[{5\over 2}]}\right){2\over {{(2\pi)}^{{5\over 2}}}}\sum_{i, j=1}^2
\sum_{n=1}^{\infty}
\left({m\over {nL}}\right)^{5\over 2}K_{5\over 2}(mLn)
\cos\left(n(\theta_i - \theta_j)\right).
\label{effpota}
\end{eqnarray}
Here $m$ stands for a gauge invariant bare mass for the adjoint 
fermion. The first line
in Eq.(\ref{effpota}) comes from the gauge and ghost fields and 
the second line is the contribution from the $N_f$ massive adjoint 
fermions. The function $K_{5\over 2}(y)$ is the modified Bessel 
function and is expressed in terms of the elementary function,
%%%%%
\begin{equation}
K_{5\over 2}(y)=\left({\pi\over {2y}}\right)^{\half}
\left(1+{3\over y}+ {3\over y^2}\right)\e^{-y}.
\label{element}
\end{equation} 
Let us note that the effective potential (\ref{effpota}) becomes identical
to the one with $N_f$ massless adjoint fermions when we take the limit,
%%%%%%%%%
\begin{equation}
\lim_{m\rightarrow 0}m^{5\over 2}K_{5\over 2}(mLn)=
{{{2^{3\over 2}\Gamma\left({5\over 2}\right)}}\over {{{(nL)}^{5\over 2}}}}.
\label{limit}
\end{equation}
%%%%%%%%%%
By defining $z\equiv mL$ and noting that $\Gamma(5/2)=3\sqrt{\pi}/4$.
the effective potential is recast as
%%%%%%%%%%%
\begin{equation}
V_{eff}={3\over{4\pi^2}}{1\over L^5}
\sum_{n=1}^{\infty}{1\over n^5}
\Biggl(-3 +4N_f (1+zn+{{(zn)^2}\over 3})\e^{-zn} \Biggr) 
\times2(1+\cos(2n\theta)),
\label{effpotb}
\end{equation}
where we have used Eq. (\ref{element}).
\par
%%%%%%%%%%%%%%%
In order to see how the massive fermion affects the vacuum structure of the
model, let us first consider asymptotic behaviors of the effective potential 
with respect to $z$. If $z$ is large enough, the fermion contribution 
to the effective potential is suppressed due to the
Boltzmann like factor $\e^{-zn}$ in Eq. (\ref{effpotb}). This is 
consistent with the observation stated in the introduction that the Hosotani 
mechanism is governed infrared physics. Then, the dominant 
contribution to the potential comes from the gauge sector 
in the model. Therefore, the vacuum configuration is given, in 
this limit, by $\theta=0~(\mbox{mod}~\pi)$~\cite{hosotanib}. 
The $SU(2)$ gauge symmetry is not broken in the 
limit\footnote{The two configurations $\theta=0$ and $\theta=\pi$ are 
physically equivalent and is related by $Z_2$ symmetry.}. 
On the other hand, if we take the massless limit, $z\rightarrow 0$, the 
effective potential, as we have mentioned above, becomes
identical to the one for the case of $N_f$ massless adjoint fermion. It has 
been known that the vacuum configuration for this case is given by
$\theta=\pi/2$~\cite{davies}. The $SU(2)$ gauge symmetry is
spontaneously broken down to $U(1)$ in this limit. 
The above observations strongly suggest that there must exist 
certain critical values of $z$, at which the gauge 
symmetry breaking patterns change.
\par
%%%%%%%%%%
In order to confirm the existence of the expected critical 
values of $z$, let us 
study the stability of the configuration $\theta=0~(\pi/2)$ with 
respect to $z$, which corresponds to the vacuum configuration 
in limit of $z\rightarrow \infty~(0)$. And we set $N_f=1$ for simplicity. 
By simple numerical calculations, we find that 
the sign of the second derivative of the effective potential 
at $\theta=\pi/2~(0)$ changes at $z\simeq 1.54135~(1.13501)$ and becomes 
negative (positive) for larger values of it. 
Both configurations are stable for $1.13501~\rough{<}~z~\rough{<}~1.54135$. 
If we compare the potential energy for the two configurations in the
narrow region, we find that $
V_{eff}(\theta={\pi/2})< (>) V_{eff}(\theta=0)~\mbox{for}~
z < (>)~z_*\simeq 1.40087$. 
\par
%%%%%%%%%%%%%%%%
These observations mean that the gauge symmetry breaking patterns 
change at $z_*$. As far as our numerical analyses are concerned, there 
appears two degenerate minima at $z_*\simeq 1.40087$ and is finite height of 
potential barrier between the configuration $\theta=0$ and $\theta=\pi/2$. 
If $z$ becomes smaller than $z_*$, the minimum of the effective potential 
locates at $\theta=\pi/2$, while if $z$ becomes larger than $z_*$, then, 
the vacuum configuration is given by $\theta=0~(\mbox{mod}~\pi)$. Hence, we 
conclude that\footnote{The phase transition at $z_*$ is the first 
order.}   
%%%%%%%
\begin{equation}
\mbox {gauge symmetry breaking pattern}=\left\{\begin{array}{lll}
SU(2)\rightarrow U(1) &\mbox{for}&  z < z_*,\\[0.3cm]
SU(2)\rightarrow SU(2)  & \mbox{for}& z > z_*.
\end{array}\right.
\label{patterna}
\end{equation}
\par
%%%%%%%%%%%
We have also computed the critical values of $z$ in case $N_f=3, 6, 10$.
They are given by $z_* \simeq 3.55154, 4.61892, 5.3521$, respectively. If we 
take $N_f=100$, the critical value is about $z_*\simeq 8.378$.
\par
%%%%%%%%%%
Let us comment on the adjoint Higgs scalar, which is originally the component
gauge field for the $S^1$ direction. The mass term for the 
scalar, which is massless at the tree level, is generated through the quantum 
correction in the extra dimensions~\cite{hosotani}. The mass term is given by
estimating the second derivative of the effective 
potential (\ref{effpotb}) at the absolute minimum. Our numerical analyses 
tell us that the Higgs scalar is always massive for each gauge
symmetry breaking pattern in Eq.(\ref{patterna}).
%%%%%%%%%%%%%%
%%%%%%%%%%%%%%
\section{Supersymmetric gauge model}
In this section let us study the ${\cal N}=1$ supersymmetric $SU(2)$ gauge model
with $N_f$ adjoint hypermultiplets defined on $M^4\otimes S^1$. The bare mass
terms, which are gauge invariant, are introduced in such a way
that they break the supersymmetry. 
Here we are interested in how the gauge symmetry breaking patterns are 
modified by such the supersymmetry breaking terms and comparing results
with those by the model with the Scherk-Schwarz mechanism of 
supersymmetry breaking. 
\par
%%%%%%%%%%%%%
In the model we have a five-dimensional 
vectormultiplet ${\cal V}=(A_{\hmu}, \Sigma, \lambda_D)$, where
$A_{\hmu}$ is the five-dimensional gauge potential and 
$\lambda_D (\Sigma)$ stands for a Dirac 
spinor\footnote{Let us note that $n$ Dirac spinors are equivalent to $2n$ 
symplectic (pseudo) Majorana spinors. $\lambda_D$ can be decomposed into two 
Majorana spinors $\lambda, \lambda'$ in four dimensions.} (a real scalar). 
Here we call $\lambda_D$ gaugino. We also have $N_f$ adjoint hypermultiplets 
${\cal H}=(\psi_D, \phi_i)$, where $\psi_D$ is a Dirac 
spinor and $\phi_{i(=1,2)}$ is a 
complex scalar called squark. In case of the Scherk-Schwarz mechanism 
of supersymmetry breaking, the gaugino and squark 
masses are shifted by an unique 
nontrivial phase associated with the $SU(2)_R$ symmetry, so that 
they have different mass terms from their superpartners in four 
dimensions. The supersymmetry is broken by the unique phase. 
Here instead of resorting to the Scherk-Schwarz 
mechanism, we add the gauge invariant bare mass for the 
gaugino $(m_g)$ and squarks $(m_s)$, as one of the 
examples, to break the supersymmetry.
\par
%%%%%%%%
Following the standard procedure to calculate the effective potential 
for the constant background gauge 
field $gL\vev{A_y}=\mbox{diag}(\theta_1, \theta_2)
=\mbox{diag}(\theta, -\theta)$, we obtain 
that\footnote{We have ignore quantum corrections to the vacuum expectation 
values for the squark fields for simplicity.}     
%%%%%%%%%
\begin{eqnarray}
V_{eff}&=&(-3-1){{\Gamma({5\over 2})}\over {\pi^{5\over 2}}}{1\over L^5}
\sum_{i,j=1}^2\sum_{n=1}^{\infty}{1\over n^5}\cos(n(\theta_i -\theta_j))
\nonumber\\
&+&{2\over{(2\pi)^{5\over 2}}}( 2^{[{5\over 2}]})
\sum_{i,j=1}^{2}\sum_{n=1}^{\infty}
\left({{m_g}\over {n L}}\right)^{5\over 2}
K_{5\over 2}(z_g n)\cos(n(\theta_i -\theta_j))
\nonumber\\
&-&{{2}\over{(2\pi)^{5\over 2}}}(4N_f)\sum_{i,j=1}^2\sum_{n=1}^{\infty}
\left({{m_s}\over {nL}}\right)^{5\over 2}
K_{5\over 2}(z_s n)\cos(n(\theta_i -\theta_j))
\nonumber\\
&+&{{\Gamma({5\over 2})}\over{\pi^{5\over 2}}}(2^{[{5\over 2}]}N_f){1\over L^5}
\sum_{i,j=1}^{2}\sum_{n=1}^{\infty}{1\over n^5}\cos(n(\theta_i -\theta_j)),
\label{effpotc}
\end{eqnarray}
where we have defined $z_g\equiv m_gL, z_s\equiv m_sL$. The first and second 
lines in Eq.(\ref{effpotc}) come from the vectormultiplet. The third 
and fourth lines are the contributions from the $N_f$ adjoint 
hypermultiplets. Here we 
have assumed that the two complex scalars in the 
hypermultiplet have a common bare mass $m_s$. The bare mass $m_g (m_s)$
explicitly breaks the supersymmetry as it should. In fact, if we take the 
limit of $m_g(m_s)\rightarrow 0$ and utilizing (\ref{limit}), the effective 
potential (\ref{effpotc}) vanish and the original ${\cal N}=1$ supersymmetry 
in five dimensions is restored.
\par
%%%%%%%%%%%%%%%% 
The effective potential is recast, by using Eq.(\ref{element}), as
%%%%%%%%%%%%%
\begin{equation}
V_{eff}=4\left({3\over{4\pi^2}}\right)
{1\over L^5}\sum_{i,j=1}^2\sum_{n=1}^{\infty}
{1\over n^5}\left(N_f F(n,z_s)-F(n,z_g)\right)\times 2(1+\cos(2n\theta)), 
\label{effpotd}
\end{equation}
where we have defined 
%%%%%%%
\begin{equation}
F(n, z_i)\equiv 1-(1+nz_i + {{(nz_i)^2}\over{3}})\e^{-nz_i}, \qquad i=g, s. 
\end{equation}
The function $F(n, z_i)$ satisfies $0\leq F(n, z_i)\leq 1$, where the 
first (second) equality holds when $z_i\rightarrow 0~(\infty)$. 
\par
%%%%%%%%%% 
Let us first study the case of $z_g=z_s(\equiv z_c)$. The effective 
potential (\ref{effpotd}) becomes 
%%%%%%%%%
\begin{equation}
V_{eff}=4\left({3\over{4\pi^2}}\right)
{{N_f-1}\over L^5}\sum_{i,j=1}^2\sum_{n=1}^{\infty}
{1\over n^5}F(n,z_c)\times 2(1+\cos(2n\theta)).
\label{effpotf}
\end{equation}
It is easy to see that the potential vanishes for $N_f=1$. This is because one  
can still have ${\cal N}=1$ supersymmetry by recombining the two 
massive fields into the same multiplet, so that there is one massless 
and one massive multiplet. And each multiplet is supersymmetric under 
the ${\cal N}=1$ supersymmetry. As a result, the total action is invariant 
under the supersymmetry. 
\par
%%%%%%%%%%%%%%%
The nonvanishing potential is 
given for $N_f \geq 2$. The supersymmetry is 
broken by an unique parameter $z_c$ in this case.
Taking $0\leq F(n, z) \leq 1$ into account, the minimum of 
the potential is always located at $\theta=\pi/2$, independent of the values 
of $z_c(\neq 0)$. Thus, the $SU(2)$ gauge symmetry is 
broken to $U(1)$. This is the same 
result as the one obtained by the Scherk-Schwarz mechanism, in which the 
unique nontrivial phase associated with $SU(2)_R$ symmetry breaks 
the supersymmetry, and the vacuum configuration is given 
by $\theta=\pi/2$~\cite{takenaga}\footnote{The effective potential for the 
case of the Scherk-Schwarz mechanism is given by
%%%%%%%%
$$
V_{eff}^{SS}={{\Gamma{(5/2)}}\over{\pi^{5/2}L^5}}(4N_f-4)
\sum_{n=1}^{\infty}{1\over n^5}
[1-\cos(n\beta)]\left(2+2\cos(2n\theta)\right).
$$
The Hosotani mechanism depends only on matter contents, so that we 
can quote the results obtained in~\cite{takenaga}.}. Let us note that 
the adjoint Higgs scalar in this case is always massive 
except for $z_c=0$, where the potential vanishes due to the 
original ${\cal N}=1$ supersymmetry in five dimensions. 
\par
%%%%%%%%%%%%%%%%%
Let us next consider the case $z_g\neq z_s$ with $N_f=1$. 
The sign of a function $C\equiv F(n, z_s)-F(n, z_g)$ is important to 
determine the minimum of the effective potential (\ref{effpotd}). 
For $z_s < (>) z_g$, the sign of $C$ is negative (positive), so that the 
configuration $\theta=0, (\pi/2)$ is realized as the vacuum 
configuration. Therefore, depending on the relative magnitude between 
$z_g$ and $z_s$, the vacuum configuration is different and accordingly, the 
gauge symmetry breaking patterns are different. If $z_g=z_s$, the effective 
potential vanishes due to the survived supersymmetry explained above.
The adjoint Higgs scalar is always massive in this case.
\par
%%%%%%%%%%%
Let us finally study the case $N_f = 2$ with $z_g\neq z_s$. 
In order to demonstrate the possible effects of the bare mass
on the gauge symmetry breaking patterns, we take  
$z_g$ to be $0.1, 1.0, 10$ as an example. For each 
value of $z_g$ we study the behavior of the effective potential 
with respect to $z_s$ and find the minimum of the potential. 
We examine the stability of the configuration
$\theta=0~(\mbox{mod}~\pi)$ and $\theta=\pi/2$ with respect to $z_s$ for
the given values of $z_g$ by studying the second derivative of the effective
potential. 
\par
%%%%%%%%%%%%%
In case $z_g=0.1$, simple numerical calculations show 
that the configuration $\theta=0$ becomes unstable 
for $z_s~\rough{>}~0.0672937\equiv z_{s1}$, on the other 
hand $\theta=\pi/2$ becomes stable for 
$z_s~\rough{>}~0.0706947\equiv z_{s2}$. The configuration that minimizes the 
effective potential in the narrow region between $z_{s1}$ and $z_{s2}$ 
is still given by the configuration which breaks the $SU(2)$ gauge symmetry
to $U(1)$, though it is not $\theta =\pi/2$.
We confirm that by numerical analyses, the behavior of $\theta$ that
minimizes the potential in the
narrow region is that the $\theta$ increases gradually from zero at  
$z_{s1}$ and approaches to $\pi/2$ at $z_{s2}$. Thus, we have shown that 
the bare mass terms for the gaugino and squark can affect the gauge 
symmetry breaking patterns through the Hosotani mechanism and 
the phase transition occurs at $z_s=z_{s1}$ for $z_g=0.1$. Hence, we 
obtain that\footnote{The phase transition is the second order unlike 
the case of the nonsupersymmetric gauge model studied in the section $2$.}
%%%%%%%%%
\begin{equation}
\mbox {gauge symmetry breaking pattern}=\left\{\begin{array}{lll}
SU(2)\rightarrow SU(2) &\mbox{for}&  z_s < z_{s1},\\[0.3cm]
SU(2)\rightarrow U(1)  & \mbox{for}& z_s > z_{s1}.
\end{array}\right.
\end{equation}
It should be noted that the very small values of $\theta$, which is 
usually of order $O(1)$, is possible in this case.
This may affect mass spectrum in four dimensions. We will discuss 
this point in the last section.
\par
%%%%%%%%%%%%%%
We repeat the same analyses as above for the case $z_g=1.0~(10)$ with $N_f=2$.
The configuration $\theta=0$ becomes unstable 
for $z_s~\rough{>}~0.618288~(2.03287)\equiv z_{s1}$, while 
the configuration $\theta=\pi/2$ becomes stable 
for $z_s~\rough{>}~0.691531~(2.47766) \equiv z_{s2}$. 
The qualitative behavior of $\theta$ that minimizes the effective potential in the
narrow region is the same as that in the case $z_g=0.1$.  
\par
%%%%%%%%%%%
The adjoint Higgs scalar in this case can be massless unlike the 
previous cases. The second derivative of the effective potential 
evaluated at $\theta=0~(\pi/2)$ 
vanishes for $z_{s1}=0.0672937~(0.0706947)$. Hence, the massless state of
the Higgs scalar is possible for the fine tuned values of $z_s$. In the 
other cases $z_g=1.0, 10$, we also have massless state of the adjoint 
Higgs scalar at the values of $z_s$, where the second derivative of the potential 
evaluated at $\theta=0,~\pi/2$ vanishes.  
\par
%%%%%%%
We have seen that the gauge symmetry breaking patterns change
due to the existence of the bare mass terms for the gaugino and squark in 
the model. The $SU(2)$ gauge symmetry is not broken 
for  $z_s < z_{s1}$, on the other hand, $SU(2)$ is broken to $U(1)$ 
for $z_s > z_{s1}$ for fixed values of $z_g$ in our examples.
\par
%%%%%%%%%%
If we add a bare mass term for the Dirac spinor $\psi_D$ in the hypermultiplet
instead of the squark $\phi_i$, the structure of the effective potential 
is different from Eq.(\ref{effpotc}). It is easy to see that the $SU(2)$ gauge
symmetry is never broken for any nonzero values of the bare masses.
%%%%%%%%%%
\section{Conclusions and discussions}
We have demonstrated that the existence of the bare mass  
affects the gauge symmetry breaking patterns
through the Hosotani mechanism. We have explicitly shown that in the 
nonsupersymmetric $SU(2)$ gauge model with the massive adjoint fermions
defined on $M^4\otimes S^1$, there exist the critical 
values for $z\equiv mL$, above (below) which the 
$SU(2)$ gauge symmetry is unbroken (broken). The phase transition is 
the first order. The asymptotic behavior 
of the effective potential with respect to $z$ also suggests the
existence of the critical values of $z\equiv mL$: If the adjoint
fermion is heavy enough, corresponding to $z\rightarrow \infty$, it 
decouples from the effective potential 
and the gauge sector of the model dominates the potential. 
Hence, the $SU(2)$ gauge symmetry is not broken through the Hosotani mechanism. 
On the other hand, if we take the massless limit of the 
fermion, that is, $z\rightarrow 0$, the vacuum configuration breaks the 
$SU(2)$ gauge symmetry to $U(1)$~\cite{davies}.  
\par
%%%%%%%%%%%%%%
We have also studied the supersymmetric gauge model defined on $M^4 \otimes S^1$.
Instead of the Scherk-Schwarz mechanism of supersymmetry 
breaking, we have introduced the bare mass terms for the gaugino in the 
vectormultiplet and the squark in the hypermultiplet to break the supersymmetry.
When the number of the hypermultiplet $N_f$ is equal to one, the critical 
point is given by $z_g = z_s$, where the potential vanishes due to the 
${\cal N}=1$ supersymmetry. The $SU(2)$ gauge symmetry is broken 
to $U(1)$ for $z_s > z_g$, while the gauge symmetry is not 
broken for $z_s < z_g$. If $N_f \geq 2$ and $z_g = z_s$, then, the $SU(2)$ gauge 
symmetry is always broken to $U(1)$ as long as $z_c(\equiv z_g=z_s)\neq 0$. 
In this case, the supersymmetry is broken by 
an unique bare mass $z_c$. And the result is 
the same as the one obtained by the Scherk-Schwarz
mechanism of supersymmetry breaking, in which the supersymmetry breaking 
parameter is also an unique and the gauge symmetry is always broken to $U(1)$.
In these cases the adjoint Higgs scalar cannot be massless except
that the models have the accidental ${\cal N}=1$ supersymmetry.
\par
%%%%%%%%%%%%%%%%
We have considered the case $z_g\neq z_s$ for $N_f=2$. 
We have shown the possible effect of the bare masses on 
the gauge symmetry breaking patterns through the Hosotani 
mechanism. By choosing the certain values of $z_g$, we have investigated the
configuration that minimizes the effective potential according to the change 
of the values of $z_s$. And we have found the critical 
values of $z_{s1}$, above (below) which the gauge symmetry is broken (unbroken).
The phase transition in the supersymmetric model is the second order unlike 
the case of the nonsupersymmetric model. We have also found that the 
massless state of the adjoint Higgs scalar appears for the fine tuned values 
of $z_s$ in this case.
\par
%%%%%%%%%%%%%%
There are many issues that are not discussed in this paper. Let us comment 
on a few of them. In the supersymmetric gauge model discussed in the 
section $3$, it is important to determine the behavior of the order 
parameter $\theta$ with respect to $z_s$ precisely in the narrow
region between $z_{s1}$ and $z_{s2}$. As mentioned in the section, it is 
possible that the magnitude of the order parameter $\theta$ can be 
very small for (fine tuned) values of $z_s$. 
Then, if particle does not have a bare mass term, the mass square of $n=0$ 
mode in the Kaluza-Klein modes behaves 
like $(\theta/L)^2$, so that the order of the 
mass is highly reduced compared with the compactification 
scale $1/L$ at the tree level. Therefore, we expect the light 
particle in four dimensions through the Hosotani 
mechanism\footnote{Let us note that the mass square of the particle is 
usually of order $(\theta/L)^2\sim \left(O(1)/L\right)^2$ through 
compactification.}.
\par
%%%%%%%%%%
It may be interesting to study the case of massive fundamental fermion 
instead of the adjoint one. It has been known that 
the $SU(N)$ gauge symmetry is not broken for  
(supersymmetric) gauge model (with the Scherk-Schwarz mechanism) with 
massless fundamental 
fermion (matter)~\cite{sakamoto}\cite{takenaga}. If we take the limit of the 
heavy bare mass, the fundamental fermion decouples from the effective 
potential and the potential is dominated by the gauge sector alone. 
Then, there are $N$ physically equivalent vacua. On the other hand, in the 
massless limit, we expect that there is a single $SU(N)$ symmetric vacuum 
for $N=$ even and a doubly degenerate $SU(N)$ vacuum for $N=$ odd. Hence, we
expect from these
observations that there exist critical values of $mL$, at which 
a sort of phase transition, in which the number of the vacuum changes, occurs.
\par
%%%%%%%%%%%%%%%
It may be interesting to consider higher rank gauge group and study the
massive particle effect on the gauge symmetry breaking patterns. In 
particular, if we introduce the hierarchy among the 
bare masses, as we have done in the supersymmetric case, it may be   
expected to occur rich gauge symmetry breaking patterns. And it is also 
interesting to study the mass spectrum in four dimensions, taking
the smallness of $\theta$ into account, as discussed above. 
\par
%%%%%%%%%%%%% 
One can also expect the same phenomena in other extra dimensions such as 
the orbifold $S^1/Z_2$ for example. According to the lessons obtained in this
paper, the gauge symmetry breaking patterns change even in the case of 
the orbifold if particles possess bare mass terms. It is expected that 
degeneracy of equivalent classes of boundary 
conditions, which has been discovered and discussed 
recently in~\cite{haba}, may be lifted due to the effect of the bare mass. 
These problems are under investigation and will reported elsewhere.  
%%%%%%%%%%%%%%
\vskip 2cm
\begin{center}
{\bf Acknowledgments}
\end{center}  
The author would like to thank the Dublin Institute for Advanced Study
for warm hospitality where a part of the work was done 
and Professor Y. Hosotani for fruitful discussions.
%%%%%
\vskip 2cm
%%%%%
%\newpage
%%%%%% References %%%%%%%

%%%%%%%%%%%
\end{document}